\documentclass[a4paper, conference]{IEEEtran}
\IEEEoverridecommandlockouts
\usepackage{cite}
\usepackage{amsmath,amssymb,amsfonts, multirow}
\usepackage{algorithmic}
\usepackage{graphicx}
\usepackage{textcomp, booktabs}
\usepackage{xcolor}
\usepackage[ruled,vlined]{algorithm2e}
\def\BibTeX{{\rm B\kern-.05em{\sc i\kern-.025em b}\kern-.08em
    T\kern-.1667em\lower.7ex\hbox{E}\kern-.125emX}}
\begin{document}

\title{Dynamic Multi-Connectivity Activation for Ultra-Reliable and Low-Latency Communication\\
}

\author{
Nurul Huda Mahmood and
Hirley Alves\\
\fontsize{9}{9}\selectfont\itshape
6GFlagship.com, University of Oulu, Finland.\\
$~^{}$Email: \{NurulHuda.Mahmood, Hirley.Alves\}@oulu.fi\\
}

\maketitle

\begin{abstract}

Multi-connectivity (MC) with packet duplication, where the same data packet is duplicated and transmitted from multiple transmitters, is proposed in 5G New Radio as a reliability enhancement feature. However, it is found to be resource inefficient, since radio resources from more than one transmitters are required to serve a single user. Improving the performance enhancement vs. resource utilization tradeoff with MC is therefore a key design challenge. This work proposes a heuristic resource efficient latency-aware dynamic MC algorithm which activates MC selectively such that its benefits are harnessed for critical users, while minimizing the corresponding resource usage. Numerical results indicate that the proposed algorithm can deliver the outage performance gains of legacy MC schemes while requiring up to $45\%$ less resources.
\end{abstract}


\section{Introduction}
\label{sec:introduction}

\IEEEPARstart{T}{he newly} introduced fifth generation New Radio (5G NR) is the first cellular standard specifically designed to support multi-service communication~\cite{3gppTS38300}. More specifically, three different service classes are introduced, namely enhanced mobile broadband (eMBB), ultra-reliable low latency communication (URLLC) and massive machine type communication (mMTC). eMBB is an enhancement of the mobile broadband services of the current long term evolution (LTE) system, with the objective of supporting a peak data rate of $20$ gigabits per second (Gbps) for downlink and $10$ Gbps for uplink. In contrast, URLLC and mMTC are emerging service classes conceived to support non-conventional communication targeting new use cases and application scenarios~\cite{3gppTR38912}. 

URLLC targets applications with demanding reliability and latency requirements. For example, one of the more stringent URLLC design goal is $99.999\%$ reliability (i.e. $10^{-5}$ outage probability) at a one way user-plane latency of maximum \textit{one} millisecond (ms). URLLC use cases include industrial control networks in Industry 4.0 scenario, communication for intelligent transport services like autonomous vehicles, smart $X$ ($X=$ home, city, grid, etc.) and Tactile Internet~\cite{3gppTS38824}.

On the other hand, mMTC service aims to provide massive connectivity solutions for various Internet of Things (IoT) applications, primarily targeting low power, low cost, low data rate sensor nodes. The main design goals are supporting high density of devices (up to a million devices per squared kilometre), energy efficiency leading to upto $10$ years battery lifetime and efficient channel access~\cite{mahmood_rrmEmbbMmtc_vtcFall2016}.

Novel solution concepts are necessary to meet the challenging design targets of 5G NR services classes. Proposed state of the art solutions range from Physical layer (PHY) techniques to the higher layers concepts, such as~\cite{shapin_phy_iswcs2018, jian_singleOFDM_jsac2019, kim_urllcTI_proc2019, abreu_multiplexing_vtcSp19, pedersen_puncturedScheduling_vtcFall2017, azari_riskAwareRRM_comMag2019}

Reference~\cite{shapin_phy_iswcs2018} introduces an abstraction model to evaluate the reliability and latency of LTE and 5G NR. The authors then utilize this model to demonstrate that URLLC reliability and latency requirements in its basic form can be met in LTE and 5G NR systems, albeit at the cost of spectral efficiency. Reference~\cite{jian_singleOFDM_jsac2019} proposes to reduce the latency by minimizing the synchronization overhead at PHY by introducing a short one-symbol PHY preamble for critical wireless industrial communications. A collection of novel physical layer technologies targeting Tactile Internet, such as waveform multiplexing, channel code design, multiple-access scheme, synchronization, and full-duplex transmission are introduced in~\cite{kim_urllcTI_proc2019}. 

Poor spectral efficiency is a common shortcoming of many proposed URLLC solutions. Efficient multiplexing of different services has been proposed to address this limitation. Reference~\cite{abreu_multiplexing_vtcSp19} presents an approach to evaluate the supported URLLC and eMBB loads considering different multiplexing options. The authors demonstrate that overlaid transmission of  both traffic types supported by successive interference cancellation enabled receivers is spectrally more efficient than transmitting them over orthogonal resources. 
On a similar note, puncturing scheduled eMBB traffic to accommodate non-scheduled URLLC traffic in an spectral efficient manner is proposed in~\cite{pedersen_puncturedScheduling_vtcFall2017}. Efficient multiplexing of URLLC and eMBB traffic is also addressed in~\cite{azari_riskAwareRRM_comMag2019}. In this work, the authors propose distributed machine learning based solution that benefits from hybrid radio resource slicing and dynamic regulation of the required spectrum. 

Multi connectivity (MC) is an specific example of practical and low complexity higher layer URLLC solution designed to improve the reliability. MC is an extension of the dual connectivity (DC) feature introduced in LTE, which allows a UE to simultaneously send/receive data from two different base stations~\cite{3gppTS36300}. The initial goal in LTE was throughput enhancement via data split. In 5G NR, MC has been identified as a reliability enhancement solution using data duplication~\cite{3gppTS37340}, where the packet failure probability is reduced by independently transmitting the same data packet to the target UE from multiple base station. Reliability oriented MC is presented and evaluated in details in~\cite{mahmood_reliabilityMC_iswcs2018}, while an analytical framework to investigate the performance vs. resource utilization trade off with MC is presented in~\cite{mahmood_analyticalMC_wcnc2019}. 

Alongside the performance analysis, several algorithms addressing different MC aspects have been proposed in the literature. Reference~\cite{anwar_PHYabstraction_pimrc2018} presents a PHY abstraction model to enable MC and efficiently utilize radio resources by choosing appropriate modulation schemes and the number of links. Similarly, MC activation subject to URLLC constraints is heuristically optimized in~\cite{rao_DCalgorithm_wcnc2018}. 

This work extends the existing literature on MC by proposing a dynamic MC activation algorithm for reliability-oriented MC. As opposed to the existing MC activation algorithms proposed in the literature, both reliability and latency constraints are specifically considered in the proposed MC activation framework. More specifically, we propose to reap the reliability enhancement accorded by MC while limiting its increased resource usage by proposing an MC activation algorithm that considers the latency budget as one of the MC activation parameters. 

\textit{Organization:} For the sake of completeness, the status of MC in light of 5G NR standardization activities and the reliability/latency vs. resource usage tradeoff with MC is discussed in Section~\ref{sec:MC_overview_3gpp}. Section~\ref{sec:systemModel} describes the considered system model, outlines the design problem with the resource efficiency as a constraint, and presents the proposed resource efficient dynamic multi-connectivity algorithm. Detailed system level simulation results validating the proposed algorithm are presented in Section~\ref{sec:results}. Finally, Section~\ref{sec:conclusion} highlights the key take away messages.


\section{Overview of Multi-Connectivity in 3GPP}
\label{sec:MC_overview_3gpp}


MC in 5G NR is inherited from the DC concept in LTE. It allows a user equipment (UE) to simultaneously send/receive data from two different evolved nodeBs (eNB). The data split occurs at the packet data convergence protocol (PDCP) layer of the transmitting eNBs. At the receiving node, the transmissions from the different eNBs are individually decoded at the lower layers and then combined at the PDCP layer of the receiver, resulting in a boost of the end-user throughput~\cite{rosa_dc_comMag2016}. 

Alongside throughput enhancement, 5G NR extends LTE DC towards improved reliability by using data duplication. Instead of splitting a data flow at the PDCP layers, data duplication allows the same data packet to be transmitted independently through different eNBs, thereby resulting in a reliability boost. The different nodes are known as master node (MN) and secondary node (SN), respectively, and are interconnected via the X2/Xn interface. 


\subsection{Multi-RAT Dual Connectivity}

To enable a faster introduction of 5G NR, initial 5G NR deployments are non-standalone and complementary to LTE, reusing the existing LTE evolved packet core (EPC) as the core network. For that purpose, 3GPP has generalized the LTE DC design to enable the support of multi radio access technology DC (MR-DC), i.e., DC between NR and LTE in the downlink and uplink~\cite{3gppTS37340}. 

The control interface to the core network is established by the MN, but the radio resource control (RRC) connection to the UE is from both MN and SN. Since the connection to the MN exists before MC is setup, the MN's RRC connection pre-exists the MC setup. To configure MC and facilitate the selection of the most suitable SN, the MN instructs the UE to make channel measurements and report back the detected cells through the RRC connection.

MC setup is initiated by the MN. Once suitable SNs are selected, the MN instructs the SN to allocate the necessary resources for MC, and to facilitate connection establishment with the target UE. If the MC setup request can be accommodated by the SN, it allocates the required resources and sends a positive acknowledgement to the MN. After the MC setup, the SN RRC further enhances control link reliability by conveying signaling to the UE through either of the nodes. 

As opposed to the control plane, the user-plane interface to the core network can be established by either of the nodes. However, the data to be transmitted to the user is transferred from the core network to only one of the nodes.


\subsection{Reliability-Oriented Multi Connectivity}

Reliability-oriented MC utilizes the multiple connections to improve reliability and, consequently, reduce the packet latency. However, it should be mentioned that it cannot impact the minimum latency, which depends on the fastest node. 

Once reliability-oriented MC is set up, the node which receives the data from the core network (known as the PDCP anchor node) duplicates the incoming data destined for the user and forwards it to the other node via the X2-U/Xn-U interface. The MN and the SN independently schedule the duplicated packets. Hence, the number of allocated radio resources and the modulation and coding scheme of the transmissions through both the MN and the SN are not necessarily the same. Moreover, the acknowledgement of the physical transmissions and the associated hybrid automatic repeat request (HARQ) mechanisms are also independent for the two links. 

\begin{figure}[htb]
    \centering
    \includegraphics[width=0.95\columnwidth]{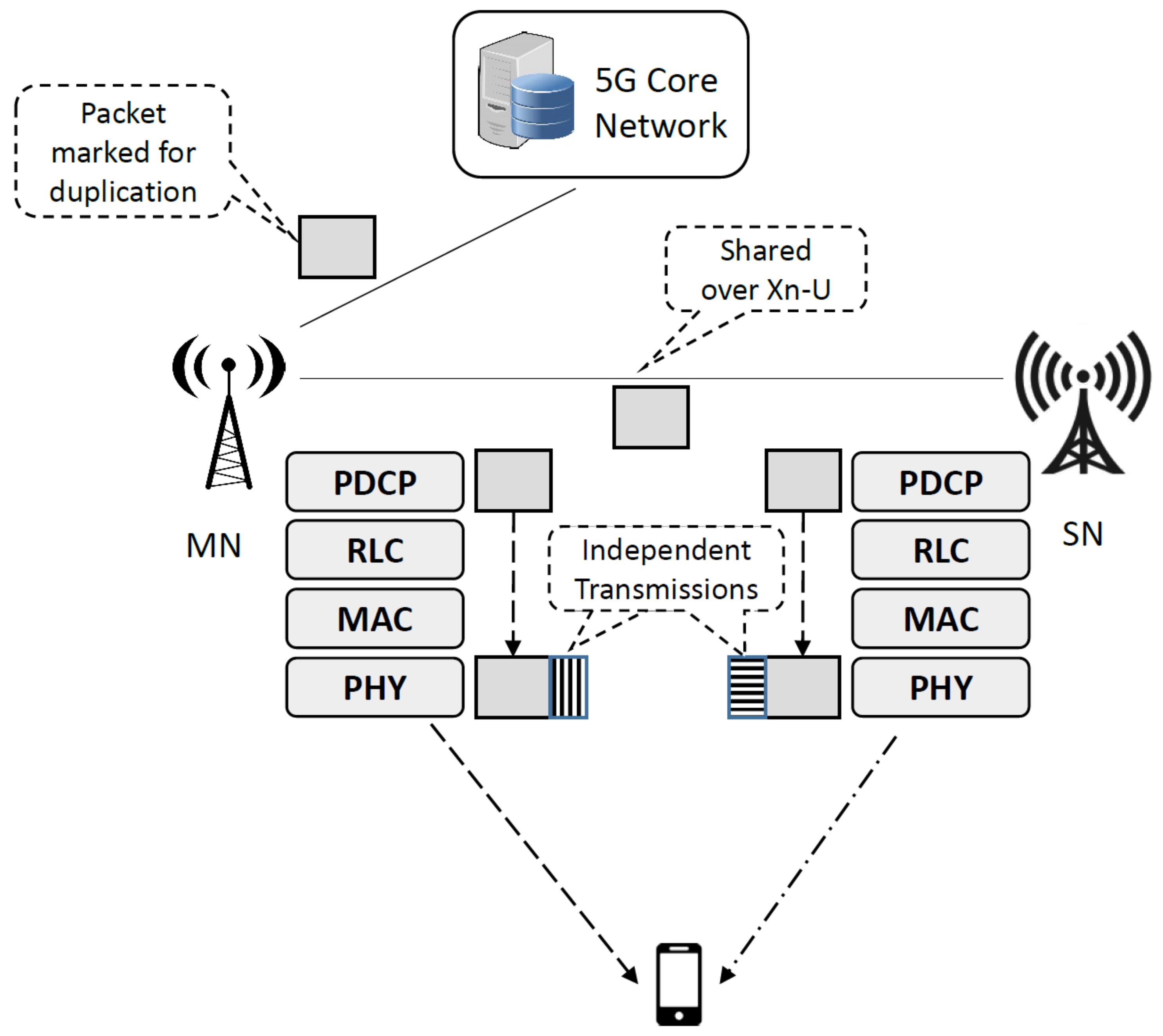}
    \caption{Schematic of downlink reliability-oriented MC in a MR-DC scenario.}
    \label{fig:NR_reliabilityDC_schematic}
\end{figure}

The UE decodes the independent transmissions separately and forwards it to the upper layers at the receiver side. The duplication of the packets are revealed at the PDCP layer, where the first successfully received packet is kept while any duplicated copies are discarded.  The operation of MC with data duplication in 5G NR downlink transmission is shown in Fig.~\ref{fig:NR_reliabilityDC_schematic}. We consider a MR-DC scenario with a LTE eNB as the MN and 5G NR next generation node B (gNB).

The user-plane protocol specifications for MC in 5G NR Release-15 is still limited to two cells. MC extending to more than two nodes is likely to be enabled in Release-16, along with the option to connect to dedicated 5G NR core network and other functionality improvements.

\section{System Model and Assumptions}
\label{sec:systemModel}

We consider a heterogeneous network scenario consisting of a macro cell and a small cell, separated by an inter-site distance of $500$ m. Intra-frequency MC is considered, i.e., the macro and the small cell are assumed to operate at different frequency bands. The downlink transmission direction is considered. 

We assume a slotted communication system. At the beginning of each slot, we assume $N$ URLLC users are randomly distributed across the entire coverage area of the macro cell. At the beginning of each time slot, each of the $N$ users can become active independently with probability $\rho < 1$. The received signal at a given URLLC user is subject to distance dependent path loss and independent and identically distributed Rayleigh fading. The transmit power is chosen such that a mean signal to noise ratio (SNR) of $3$ dB is achieved at the cell edge.

The best gain with MC is observed for users where the received signal from the MN and the SN are at similar level of strength~\cite{mahmood_reliabilityMC_iswcs2018}. In practice, this roughly corresponds to users at the cell edge. In light of this finding, MC is only activated for users whose difference in the received SNR from the two cell are within a given range denoted by $\Delta_{MC}$.


\subsection{Resource Allocation}
In order to meet the stringent latency target of URLLC services, 5G NR introduces a flexible frame structure with the option to have mini-slots of duration significantly shorter than one ms~\cite{3gppTS38211}. In this work, we consider mini-slots of duration $0.125$ ms. This allows sufficient time budget for HARQ retransmissions, even when considering the tightest latency budget. In addition, each URLLC transmission is assumed to have an corresponding latency budget, and the transmission is said to be in outage if the packet is not successfully received within this time-frame. 

We assume that the metadata (i.e., the control information needed to decode the transmission) and the data of the $l^{th}$ transmission are encoded collectively with a target block error rates (BLER) given by $P^{l}_{e}$. Chase combining, which results in an SNR boost, is assumed for retransmitted packets~\cite{FPD01_chaseCombining}. 

Considering the short packet size of URLLC traffic, the number of allocated resources for each user follows from the finite blocklength theory~\cite{polyanskiy_trIT2010}. The received SNR $\gamma_{l}$ of user $l$ is dictated by its location and the fading channel, which is assumed to known with the help of channel quality indicator (CQI) feedback. The number of information bits $L$ that can be transmitted with decoding error probability $P_{e}^{l}$ in $R_l$ channel uses in an additive white Gaussian noise channel with a given SNR $\gamma_l$ is~\cite{polyanskiy_trIT2010}
\begin{align}
	\label{eq:polyanskiy}
	L = R_l C(\gamma_l) - Q^{-1}(P_e^{l})\sqrt{R_l V(\gamma_l)} + 
		\mathcal{O}(\log_2 R_l),
\end{align}
where $C(\gamma_l) = \log_2 (1 + \gamma_l)$ is the Shannon capacity of AWGN channels under infinite blocklength regime, $V(\gamma_l) = \frac{1}{\ln(2)^2} \left( 1 - \frac{1}{\left( 1 + \gamma_l \right)^2}\right)$ is the channel dispersion (measured in squared information units per channel use) and $Q^{-1}(\cdot)$ is the inverse of the Q-function. Using the above and assuming a packet size (combined metadata and data size) of $32$ bytes, the channel usage $R_l$ can be approximated as~\cite{AV_jsac2018}
\begin{multline}
	\label{eq:channelUsageAnand}
	R_l \approx \frac{L}{C(\gamma_l)} + \frac{Q^{-1}(P_e)^2 V(\gamma_l)}{2 C(\gamma_l)^2} \times \\
	\left[ 1 + \sqrt{1 + \frac{4 L C(\gamma_l)}{Q^{-1}(P_e)^2 V(\gamma_l)}}\right].	
\end{multline}
Sufficient radio resource to accommodate all the active URLLC users at each time slot is assumed available. 

\subsection{Multi Connectivity Configuration}
MC is activated for users with a received SNR from the macro and the small cell within $\Delta_{MC}$ of each other. Following the well known selection diversity concept, the received SNR with MC is given as
\begin{align}
	\label{eq:DC_snr}
	\gamma_l = \max\left(\gamma_{l}^{macro}, \gamma_{l}^{small}\right), 
\end{align}
where $\gamma_{l}^{macro}$ and $\gamma_{l}^{small}$ are the received SNRs from the macro and the small cell, respectively. Consequently, the resulting outage probability can be derived from~\eqref{eq:polyanskiy} as 
\begin{align}
	\label{eq:outageProbCalc}
	P_e^{l} = Q\left( \frac{R_l C(\gamma_l) - L}{\sqrt{R_l V(\gamma_l)}}\right). 
\end{align}

\subsection{Proposed Algorithm}
\label{sec:algorithm}

The standard MC configuration policy states that all users fulfilling MC activation criteria should operate in MC mode. This is resource inefficient since the radio resources from all nodes are utilized to serve a single URLLC user, even when the transmission from a single node is successful. In this work, we propose to overcome this limitation by presenting a heuristic latency-aware MC configuration algorithm as outlined in Algorithm~\ref{alg:proposedAlgorithm}. The key idea is as follows: in addition to the the MC configuration parametrized by $\Delta_{MC},$ MC is only activated for users with a latency budget below a critical threshold given by $\tau$. The additional control signalling required to share the latency budget among the participating base stations is minimal with respect to conventional MC operation.

\subsubsection*{Outage Probability Calculations}
Failure in successful reception of an individual transmission attempt is determined by the outage probability $P_e^l$. For users operating in single connectivity (SC) mode, $P_e^l$ is given by the initial BLER target for the first transmission. For each subsequent retransmissions, it is given by~\eqref{eq:outageProbCalc}, where $\gamma_l$ is the Chase combined SNR of the multiple transmissions. MC leads to an SNR boost as given by~\eqref{eq:DC_snr}, which in turn leads to a lower outage probability than the BLER target even for the first transmission. 

In the event of a failure in successful reception, the respective user is rescheduled after $t^{RTT}$ time slots, corresponding to the HARQ round trip time. The latency budget is correspondingly reduced. A user is considered to be in outage if it fails to successfully transmit a packet within its latency budget.

\begin{algorithm}
\label{alg:proposedAlgorithm}
\caption{Proposed latency aware dynamic MC activation algorithm}
\SetAlgoLined
 \For{Each time slot $t$}{
 	Generate random number of URLLC users at random location and a given latency budget $d_l$\;
 	Add to transmission buffer $b_t$\;
 	\For{Each user $l$ in buffer $b_t$}{
 		SNR $\gamma_l$ determined by location and random fading\;
 		Allocate $R_l$ using~\eqref{eq:channelUsageAnand}\;
 		\If{Retransmission}{
 			Update Chase combined SNR $\gamma_l$\;
 			Update outage probability $P_e^l$ using~\eqref{eq:outageProbCalc}\;
 		}
 		\If{$|\gamma_l^{macro} - \gamma_l^{small}| < \Delta_{MC}$ \textbf{AND}
 		$d_l < \tau$}{
 			\textit{MC MODE:}\\
 			Update $\gamma_l \leftarrow  \max\left(\gamma_{l}^{macro}, \gamma_{l}^{small}\right)$\;
 			Update $P_e^l$ using~\eqref{eq:outageProbCalc}\;
 		}
 		\If{Outage event}{
 			Add user $l$ to buffer $b_{t + t^{RTT}}$\;
 			Update latency budget $d_l \leftarrow  d_l - t^{RTT}$\;
 			Increment number of transmissions for user $l$\;
 		}
	}
 }
\end{algorithm}

 
\section{Numerical Results}
\label{sec:results}

This section presents numerical validation of the proposed algorithm. The considered simulation parameters are listed in Table~\ref{tab:simParameters}. Three different scenarios are considered, namely conventional SC, state-of-the-art MC, and the proposed latency aware MC algorithm. The obtained results are averaged over a million simulation runs to ensure statistical validity.

\begin{table}[]
\begin{center}

\caption{Simulation Parameters}
\label{tab:simParameters}
\begin{tabular}{lc}
\toprule
Parameter & Value \\
\midrule
Inter site distance & $500$m \\
Path loss exponent & $4$\\
Initial BLER target & $0.1$\\
TTI size & $0.125$ ms\\
HARQ round trip time & $2$ TTIs\\
Initial latency budget & $6$ TTIs\\
Critical latency threshold & $\tau = 2$ TTIs\\
MC parameter $\Delta_{MC}$ & $20$ dB\\
Number of URLLC users & $N = 10$\\
Activation probability & $\rho = 0.3$\\
\bottomrule
\end{tabular}
\end{center}
\vspace{-6mm}
\end{table} 

\subsection{Resource Utilization}

\begin{figure}[htb]
    \centering
    \includegraphics[width=0.95\columnwidth]{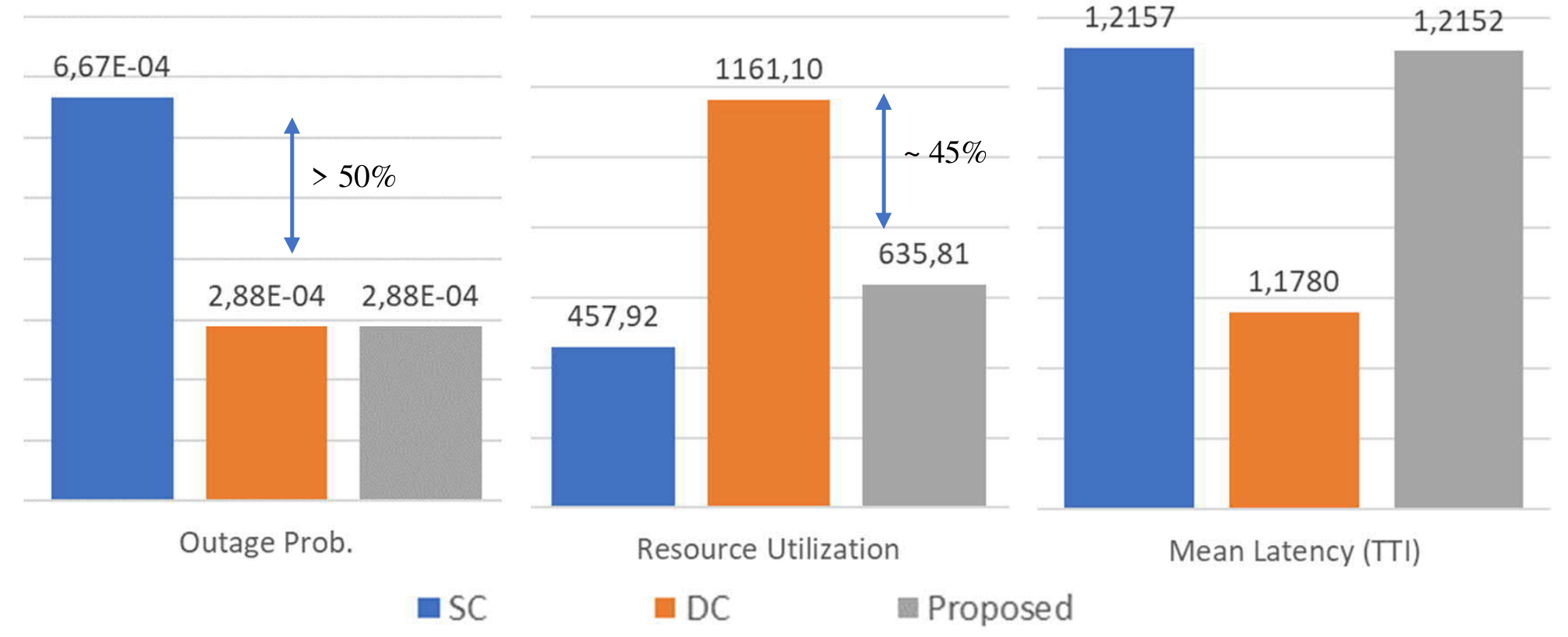}
    \caption{Key performance indicators under evaluation. An average $45\%$ gain in resource utilization is observed for the proposed algorithm with respect to MC. Notice that the proposed solution attains similar outage performance as MC, while maintaining similar average latency as the SC case.}
    \label{fig:outageProb}
\vspace{-4mm}
\end{figure}

Three different key performance indicators, namely the outage probability, the resource utilization and the mean latency are presented in Fig.~\ref{fig:outageProb}. Comparing the performance collectively, the proposed algorithm delivers the outage probability gains of MC while almost halving the required resource uses. Thus, the proposed algorithm is found to significantly enhance the resource utilization vs. performance tradeoff, a typical limitation of legacy MC.

The outage probability results reveal that the proposed algorithm's outage performance is the same as that with legacy MC. Both schemes result in halving the outage probability compared to the baseline SC. However, the proposed algorithm is able to achieve this with up to $45\%$ less resources. 

Furthermore, we observe that the mean latency of the proposed algorithm is similar to that of SC. This results from the latency-awareness of the proposed scheme. In particular, the proposed algorithm enhances resource utilization by relegating the activation of MC to users with a high probability of violating the latency constraint. Please note that though the mean latency is higher with the proposed algorithm, the latency violation probability (which is the outage probability) is the same as legacy MC. 

%

%

\subsection{Latency Results}
\begin{figure}[htb]
    \centering
    \includegraphics[width=0.99\columnwidth]{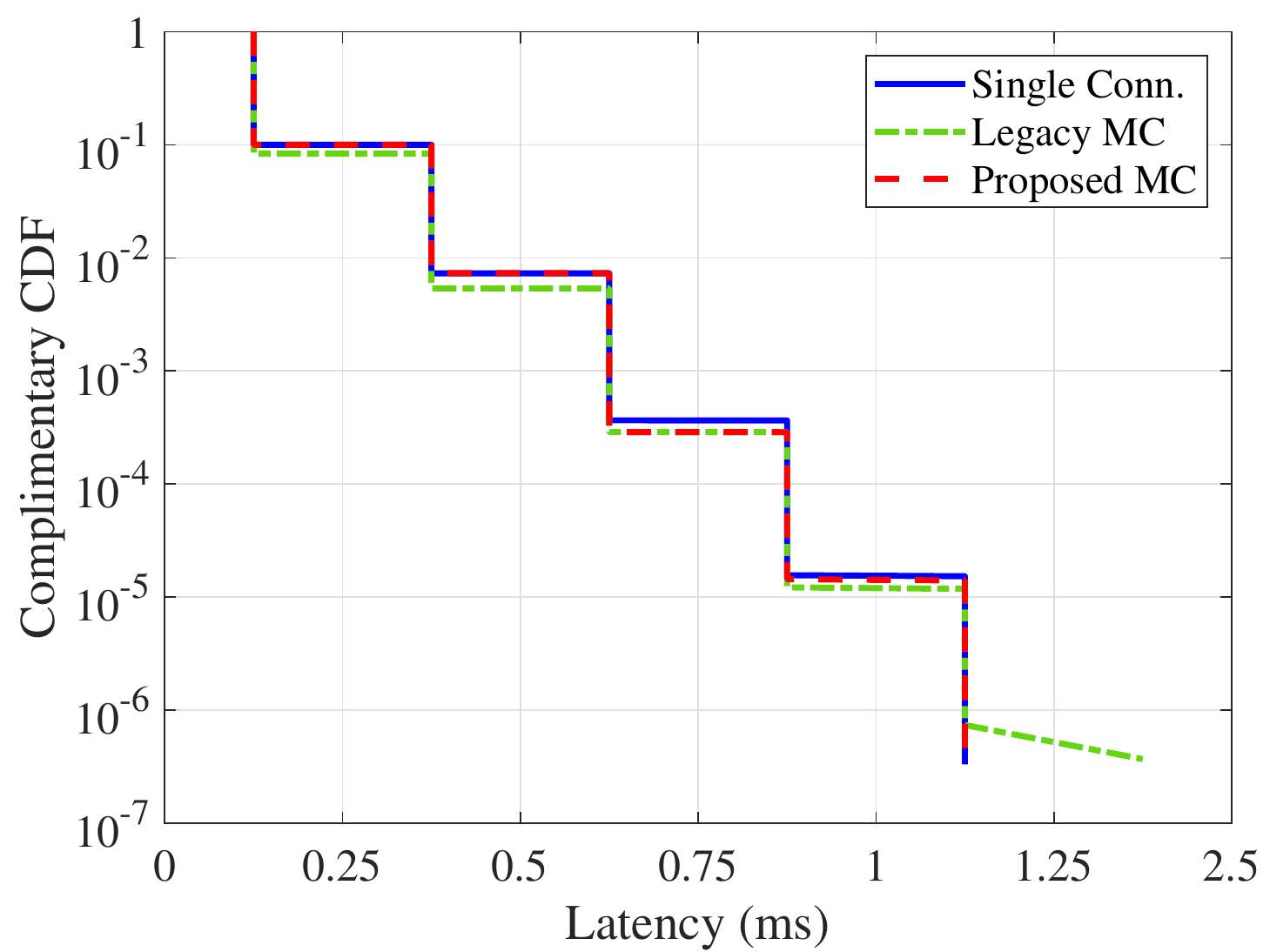}
    \caption{Latency CCDF of the entire cell area with SC, legacy MC and the proposed MC activation algorithm.}
    \label{fig:latencyCCDF}
\end{figure}

The complementary cumulative distribution function (CCDF) of the transmission latency (in ms) needed to transmit a given packet for the considered transmission strategies is shown in Fig.~\ref{fig:latencyCCDF}. The latency of all the users across the network is considered in the CCDF. The well-know staircase behaviour of the latency is observed. 

Since only a fraction of the users operate in MC mode (cf. Table~\ref{tab:MCstats}) and hence benefit from MC operation, the latency improvement with MC (both legacy and the proposed algorithm) is only marginal compared to SC mode. This conforms with the earlier findings reported in~\cite{mahmood_reliabilityMC_iswcs2018}. Moreover, legacy MC is found to result in lower latency compared to the proposed algorithm, since the proposed algorithm prioritizes resource efficiency by operating users which can accommodate additional retransmissions in SC mode. 

Table~\ref{tab:MCstats} presents statistics specifically focusing on the users operating in MC. We observe that the percentage of users in MC drops about an order of magnitude (from $36.87\%$ to $3.58\%$) with the proposed algorithm, along with a slight improvement in the outage. Thus, the additional resource usage of MC is needed for far fewer users, resulting in the significant resource utilization improvement while guaranteeing the same outage performance. The outage statistics for the same set of users in SC mode is also presented for comparison. 


\begin{table}[t]
\begin{center}

\caption{Statistics for MC users}
\label{tab:MCstats}
\begin{tabular}{l c c c}
\toprule
 & Users in MC & Outages & Avg. nr. of Tx.\\
\midrule
Proposed MC & $10,741 (3.6\%)$ & $3$ & $2.05$\\
Legacy MC & $110,210 (36.9\%)$ & $8$ & $2.21$\\
Single Conn. & $-$ & $29$ & $1.11$\\
\bottomrule
\end{tabular}
\end{center}
\vspace{-5mm}
\end{table} 
%
%
\section{Conclusions}
\label{sec:conclusion}

Multi-connectivity is proposed as a potential reliability enhancement solution for URLLC applications. However, outage probability enhancement with legacy MC schemes is resource inefficient. In order to enhance the performance improvement vs. resource utilization tradeoff typically associated with MC, this work proposes and numerically verifies a resource efficient latency-aware dynamic MC algorithm. The proposed heuristic algorithm operates by activating MC only for users with a high latency violation probability instead of all users that fulfil the MC activation criteria. It is found to deliver the outage performance gains of legacy MC while requiring up to $45\%$ less resources.

\section*{Acknowledgement}
This work has performed in the framework of Academy of Finland 6Genesis Flagship (grant no. $318927$). 

\bibliographystyle{IEEEtran}
\bibliography{DC_bibliography}

\end{document}